\RequirePackage{fix-cm}
\documentclass[smallextended]{svjour3}       
\smartqed  
\usepackage{graphicx}

\usepackage[intlimits]{amsmath}
\usepackage{graphics}
\usepackage{amsfonts}
\usepackage{amssymb}
\usepackage{dsfont}
\usepackage{units}

\usepackage[usenames]{color}
\usepackage{colortbl}
\usepackage[colorlinks=true,linkcolor=blue,citecolor=blue,urlcolor=blue]{hyperref}

\usepackage{titlesec}
\usepackage{framed}
\usepackage{eurosym}
\usepackage{bbding}
\usepackage{cite}
\usepackage{slashed}

\usepackage{bm}
\usepackage{wasysym}
\usepackage{multirow}

        \usepackage[vcentermath]{youngtab}

\def\p{\partial}
\newcommand{\conjg}[1]{\ensuremath{\hspace{1pt}\overline{\hspace{-1pt}#1\hspace{-1pt}}}\hspace{1pt}}
\newcommand{\vect}[1]{{\mbox{\boldmath $#1$}}}

\def\mA{\ensuremath{\mathcal{A}}}

\def\mD{\ensuremath{\mathcal{D}}}

\def\mL{\ensuremath{\mathcal{L}}}

\def\mS{\ensuremath{\mathcal{S}}}

\graphicspath{
	{./figures/}
	{../figures/}
}

\begin{document}

\title{Theory introduction to baryon spectroscopy \thanks{This work is supported by the FCT Investigator Grant IF/00898/2015.}
}


\author{Gernot Eichmann
}


\institute{G. Eichmann \at
              LIP Lisboa, Av. Prof. Gama Pinto 2, 1649-003 Lisboa, Portugal \\
              Instituto Superior T\'ecnico, Universidade de Lisboa, 1049-001 Lisboa, Portugal\\
              \email{gernot.eichmann@tecnico.ulisboa.pt}
}

\date{Received: date / Accepted: date}

\maketitle

\begin{abstract}
In these introductory notes I give a brief overview on some theoretical aspects of light baryon spectroscopy. 
The first part contains a discussion of the symmetries of the baryon spectrum, the construction of flavor wave functions 
and some basic features of quark models. The second part examines the need for relativistic
quantum field theory and focusses on spectrum calculations with functional methods.
\keywords{Quantum Chromodynamics \and Baryon spectroscopy \and Functional methods}
\end{abstract}

\section{Introduction} \label{sec:intro}

Even though Quantum Chromodynamics (QCD), the theory of the strong interaction,
has been established several decades ago, it
still occupies a special place within the Standard Model. One remarkable feature is that
its elementary degrees of freedom, quarks and gluons, are not observable due to color confinement.
What we measure in detectors are hadrons, like mesons as $q\bar{q}$ states and baryons with three valence quarks.  
To understand the dynamics of QCD, we must therefore investigate the spectrum and interactions of hadrons.

Light baryon spectroscopy has undergone much experimental progress in recent years~\cite{Klempt:2009pi,Tiator:2011pw,Aznauryan:2011qj,Crede:2013sze,Thiel:2022xtb}.
Its theoretical description  remains, however, challenging for several reasons.
The strong coupling between quarks and gluons
at low momenta requires nonperturbative methods.
While nonrelativistic quark models have established an efficient baseline,
light baryons feel the effects of relativity and chiral symmetry
and thus one needs quantum field theory (QFT) to describe them.
One prominent feature of QCD is
dynamical mass generation: Three current quarks  contribute only about 1\% to the mass of the proton,
so the overwhelming majority must be generated in QCD. But how?
Most hadrons are also resonances and decay into other hadrons, which complicates their dynamics substantially.
In addition, there is evidence for exotic hadrons such as tetraquarks and pentaquarks, which can mix with ordinary hadrons,
so that light baryons could have substantial multiquark components.

The purpose of these notes is to give a brief pedagogical introduction to light baryon spectroscopy from a theory point of view.
After a survey of the symmetries of the  spectrum, the construction of  flavor wave functions and some basic properties of quark models,
we discuss how baryons emerge in QFT and eventually focus on functional methods.
The material is mainly based on a  lecture course held at IST Lisbon~\cite{QCD-lns} and two reviews~\cite{Eichmann:2016yit,Barabanov:2020jvn}.

\begin{figure*}
  \centering
  \includegraphics[width=1\textwidth]{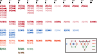}
\caption{Light and strange baryon spectrum from the PDG~\cite{ParticleDataGroup:2020ssz} up to $J^P = 11/2^+$.
         The columns correspond to different $J^P$ and the colors to isospin and hypercharge. The different font weights represent four-, three- and two-star resonances.} 
\label{fig:spectrum-1}
\end{figure*}

\section{Symmetries of the spectrum} \label{sec:symmetries}

   To begin with, let us have a look at the experimentally known baryon spectrum for states made of light ($n=u,d$) and strange ($s$) quarks, which is collected in Fig.~\ref{fig:spectrum-1}.
   The columns are arranged with respect to $J^P$, where $J$ is the total angular momentum and $P$ the parity of a given state.
   Three quarks, each with spin $\tfrac{1}{2}$, can produce a total spin $\tfrac{1}{2}$ or $\tfrac{3}{2}$,
   and combined with orbital angular momentum $L$ this gives the total angular momentum $J$ (which is usually also called `spin').
   The different colors correspond to isospin and hypercharge, and in each slot there is a ground state with further possible excited states.
   How do these quantum numbers arise in the first place?

   Since QCD describes the strong interaction, each quantum number must correspond to a symmetry of the QCD Lagrangian $\mL$, or
    its action $S$:
   \begin{equation}
      S = \int d^4x\,\mL\,, \qquad
      \mL = \conjg\psi\left(i\slashed{\p} + g\slashed{A} - \mathsf{M} \right) \psi - \frac{1}{4} F_{\mu\nu}^a F^{\mu\nu}_a\,.
   \end{equation}
   The quark fields $\psi_{\alpha,i,f}(x)$ carry a Dirac index $(\alpha = 1 \dots 4)$, a color index $(i=1\dots 3)$ and a flavor index $(f=1\dots N_f)$,
   and $\mathsf{M} = \text{diag}(m_u, m_d, m_s, \dots)$ is the quark mass matrix in flavor space.
   What are the symmetries of the action?

    {\tiny$\blacksquare$} For the spectrum, the local $SU(3)_c$ gauve invariance of the action does not tell us much apart from the fact that hadrons must be color singlets.
    When we combine three quarks, which transform under the fundamental $\mathbf{3}$ representation of $SU(3)_c$,
    baryons can thus only belong to the totally antisymmetric color-singlet representation $\mathbf{1}_A$:
    \begin{equation}\label{333}
       \mathbf{3}\otimes\mathbf{3}\otimes\mathbf{3} = \mathbf{10}_S \oplus \mathbf{8}_{M_A} \oplus \mathbf{8}_{M_S} \oplus \mathbf{1}_A\,.
    \end{equation}

    \smallskip

    {\tiny$\blacksquare$} The action is invariant under the Poincaré group, which consists of translations, rotations and boosts.
    The Poincaré group has two Casimir operators labelling the states: the mass $M$ and the total angular momentum $J$.

    \smallskip

    {\tiny$\blacksquare$} The action is invariant under parity, charge conjugation and time reversal.
    Their combination ($CPT$) is always conserved, and charge conjugation transforms baryons into antibaryons, so that only parity induces a conserved quantum number $P$.

    \smallskip

    So far we have $M$ and $J^P$ as quantum numbers labelling the states.
    There are various further global transformations of the quark fields, which 
    we exemplify for $N_f = 3$ with three flavors $u$, $d$ and $s$:

    \smallskip

    {\tiny$\blacksquare$} The group $U(1)_V$ consists of global phase transformations $\psi' = e^{i\varepsilon}\psi$
    which leave $\mL$ invariant. The conserved quantum number is the baryon number $B= \frac{1}{3}(n_u + n_d + n_s)$,
    where $n_q$ is the number of quarks minus antiquarks in a state. Baryons carry $B=1$ (and so do pentaquarks).

    \smallskip

    {\tiny$\blacksquare$} The transformation $\psi' = e^{i\varepsilon}\psi$, where $\varepsilon = \sum_{i=1}^8 \varepsilon_a \mathsf{t}_a$ and
    $\mathsf{t}_a$ are the eight generators of $SU(3)$, define the group $SU(N_f)_V$. It induces vector currents and charges:
    \begin{equation}
       V_a^\mu = \conjg\psi \gamma^\mu \mathsf{t}_a \psi\,, \qquad
       \p_\mu V_a^\mu = i\conjg\psi \left[ \mathsf{M}, \mathsf{t}_a \right] \psi\,, \qquad
       Q_a^V(t) = \int d^3x\,\psi^\dag \mathsf{t}_a \psi\,.
    \end{equation}
    This is the usual flavor symmetry, which by the Noether theorem is preserved if the divergences of the currents vanish.
    This can only work if $\mathsf{M} = m$, i.e., all quark masses are equal so that the commutator is zero.
    On the other hand,  the diagonal $SU(3)$ generators $\mathsf{t}_3$ and $\mathsf{t}_8$ commute with $\mathsf{M}$ and are  always conserved.
    The corresponding quantum numbers are the isospin-3 component $I_3$ and the hypercharge $Y$,
    \begin{equation}
       I_3 = \frac{1}{2}(n_u-n_d)\,, \qquad
       Y = \frac{1}{3} (n_u + n_d - 2n_s)\,,
    \end{equation}
    which are still good quantum numbers to label the states even if the flavor symmetry is broken. 

    \smallskip

    {\tiny$\blacksquare$} The transformation $\psi' = e^{i\gamma_5\varepsilon}\psi$, again with $\varepsilon = \sum_{i=1}^8 \varepsilon_a \mathsf{t}_a$,
    define the group $SU(N_f)_A$ which induces axialvector currents
    \begin{equation}
       A_a^\mu = \conjg\psi \gamma^\mu\gamma_5 \mathsf{t}_a \psi\,, \qquad
       \p_\mu A_a^\mu = i\conjg\psi \left\{ \mathsf{M}, \mathsf{t}_a \right\} \gamma_5 \psi
    \end{equation}
    and corresponding charges $Q_a^A(t)$. The combination $SU(N_f)_V \times SU(N_f)_A$ defines \textit{chiral symmetry}.
    The axialvector symmetry is classically conserved only in the chiral limit $\mathsf{M} = 0$, where the anticommutator vanishes.
    However, it is spontaneously broken in the QFT by the quark-gluon dynamics,
    which is called \textit{dynamical chiral symmetry breaking}. 

    \smallskip

    {\tiny$\blacksquare$} The axial $U(1)_A$ symmetry defined by $\psi' = e^{i \gamma_5 \varepsilon}\psi$, where $\varepsilon$ is again just a number,
    is classically also only conserved for $\mathsf{M}=0$. However, this symmetry does not survive the quantization of QCD and is \textit{anomalously} broken.

    \smallskip

    All in all, the quantum numbers to label baryons with three quark flavors are the mass $M$, spin and parity $J^P$, the isospin $I_3$ and the hypercharge $Y$.
    This is how we arranged the states in Fig.~\ref{fig:spectrum-1}. In other words: The symmetries of the spectrum are the symmetries of the QCD Lagrangian!

    \begin{figure}
      \centering
      \includegraphics[width=0.45\textwidth]{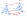}
      \caption{Cayley graph for the permutation group $S_3$.}
      \label{fig:cayley}
    \end{figure}

\section{Flavor wave functions} \label{sec:flavor-wfs}

    To construct the baryon flavor wave functions, one must find the combined irreducible representations of
    $SU(N_f)$ (for $N_f$ flavors) and the permutation group $S_3$ (for three quarks).
    There is a quick and straightforward way to do so. 
    The permutation group $S_3$ consists of six permutations, which are visualized by the Cayley graph in Fig.~\ref{fig:cayley}.
    Each permutation can be reconstructed from a transposition $P_{12}$, which exchanges $1\leftrightarrow 2$,
    and a cyclic permutation $P_{123}$, which transforms $1\to 2$, $2\to 3$ and $3\to 1$:
    \begin{equation} \renewcommand{\arraystretch}{1.2}
    \begin{array}{rl}
       1\,\psi_{123} &= \psi_{123}\,, \\
       P_{123}\,\psi_{123} &= \psi_{231}\,, \\
       P_{123}^2\,\psi_{123} &= \psi_{312}\,,
    \end{array}\qquad
    \begin{array}{rl}
       P_{12}\,\psi_{123} &= \psi_{213}\,, \\
       P_{12}\,P_{123}\,\psi_{123} &= \psi_{132}\,, \\
       P_{12}\,P_{123}^2\,\psi_{123} &= \psi_{321}\,.
    \end{array}
    \end{equation}
    The task is to find combinations of the $\psi_{ijk}$ that transform under irreducible representations of $S_3$
    and hence define invariant subspaces under permutations:

    \smallskip
    {\tiny$\blacksquare$} A singlet $\mS$ is invariant under any permutation: $P_{12} \,\mS = \mS$, $P_{123} \,\mS = \mS$.
    This is satisfied by $\mS = \psi_{123} + \psi_{231} + \psi_{312} + \psi_{213} + \psi_{132} + \psi_{321}$.

    \smallskip
    {\tiny$\blacksquare$} An antisinglet (antisymmetric singlet) $\mA$ is invariant under a cyclic permutation but antisymmetric under transpositions, $P_{12} \,\mA = -\mA$, $P_{123} \,\mA = \mA$,
    which is satisfied by $\mA = \psi_{123} + \psi_{231} + \psi_{312} - \psi_{213} - \psi_{132} - \psi_{321}$.

    \smallskip
    {\tiny$\blacksquare$} The remaining four combinations result in two doublets $\mD_1$, $\mD_2$ whose upper components are mixed-antisymmetric and lower components mixed-symmetric under an exchange $1\leftrightarrow 2$:
    \begin{equation}
    \begin{split}
       \mD_1 &= \left[ (\psi_{231} - \psi_{132}) - (\psi_{312} - \psi_{321} ) \atop \frac{1}{\sqrt{3}}\left( (\psi_{231} + \psi_{132}) + (\psi_{312} + \psi_{321} ) - 2\,(\psi_{123} + \psi_{213})\right) \right], \\[1mm]
       \mD_2 &= \left[ -\frac{1}{\sqrt{3}}\left( (\psi_{231} - \psi_{132}) + (\psi_{312} - \psi_{321} ) - 2\,(\psi_{123} - \psi_{213})\right) \atop (\psi_{231} + \psi_{132}) - (\psi_{312} + \psi_{321} )   \right].
    \end{split}
    \end{equation}
    They transform like $P_{12}\,\mD_j = \mathbf{M}_{12}^T\,\mD_j$, $P_{123}\,\mD_j = \mathbf{M}_{123}^T\,\mD_j$, where $\mathbf{M}_{12}$ and $\mathbf{M}_{123}$
    are the two-dimensional matrix representations of $S_3$:
    \begin{equation}
       \mathbf{M}_{12} = \left( \begin{array}{r@{\quad}r} -1 & 0 \\ 0 & 1 \end{array}\right), \qquad
       \mathbf{M}_{123} = \frac{1}{2}\left( \begin{array}{r@{\;\;}r} -1 & -\sqrt{3} \\ \sqrt{3} & -1 \end{array}\right).
    \end{equation}
    In this way, we have rearranged the six $\psi_{ijk}$ into permutation-group multiplets,
    which transform under irreducible representations of $S_3$.
    In terms of Young diagrams, they correspond to
        \begin{equation*}
             \scriptsize\yng(3) \;\dots \; \mS\,,  \qquad
             \scriptsize\yng(2,1)  \;\dots \;\mD_j\,, \qquad
             \scriptsize\yng(1,1,1)  \;\dots \;\mA \,.
        \end{equation*}

    \bigskip

    \begin{table}
    \centering
    \caption{Combined $S_3$ and $SU(3)$ classification for light and strange baryons.}
    \label{tab:flavor}
    \begin{tabular}{l @{\qquad\quad} llll @{\qquad\quad} lll @{\qquad\quad} ll @{\qquad\quad} l} \hline\noalign{\smallskip}
            & $uuu$         & $uud$      & $ddu$      & $ddd$      & $uus$      & $uds$       & $dds$      & $ssu$   & $ssd$   & $sss$      \\ \noalign{\smallskip}\hline\noalign{\smallskip}
    $\mS$   & $\Delta^{++}$ & $\Delta^+$ & $\Delta^0$ & $\Delta^-$ & $\Sigma^+$ & $\Sigma^0$  & $\Sigma^-$ & $\Xi^0$ & $\Xi^-$ & $\Omega^-$ \\[0.5mm]
    $\mD_1$ &               & $p$        & $n$        &            & $\Sigma^+$ & $\Sigma^0$  & $\Sigma^-$ & $\Xi^0$ & $\Xi^-$ &            \\[0.5mm]
    $\mD_2$ &               &            &            &            &            & $\Lambda^0$ &            &         &         &            \\[0.5mm]
    $\mA$   &               &            &            &            &            & $\Lambda^0$ &            &         &         &            \\ \noalign{\smallskip}\hline
    \end{tabular}
    \end{table}

    Now let us combine this with $SU(3)$. If we consider a state with quark content $uud$ and write $\psi_{123} = u_1 u_2 d_3 = (uud)_{123}$,
    $\psi_{231} = u_2 u_3 d_1 = (duu)_{123}$ and so on, and plug this into the formulas above, we find
    \begin{equation}
      \mS \propto uud + udu + duu\,, \qquad
      \mD_1 \propto  \left[ udu - duu \atop -\frac{1}{\sqrt{3}}\left( udu + duu - 2uud\right) \right],
    \end{equation}
    together with $\mA  = \mD_2 = 0$. Apart from overall normalization, $\mD_1$ is the flavor wave function of the proton
    and $\mS$ the one for the $\Delta^+$. For a quark content $ddu$ we only need to exchange $u\leftrightarrow d$, which
    gives the flavor wave functions for the neutron and the $\Delta^0$. The combination $uuu$ only returns a singlet ($\Delta^{++}$) and nothing else,
    and so does $ddd$ ($\Delta^-$). To find the flavor wave functions for $uus$ ($\Sigma^+$),  replace $d\to s$ above,
    and likewise for $dds$ ($\Sigma^-$), $ssu$ ($\Xi^0$), $ssd$ ($\Xi^-$) and $sss$ ($\Omega^-$).
    The combination $uds$ returns everything: a singlet, two doublets and an antisinglet.

    The resulting flavor wave functions for $SU(3)_f$ are collected in Table~\ref{tab:flavor}.
    Counting up the rows, we find ten singlets, eight doublets and one antisinglet of the group $S_3$ ---
    or in other words, one decuplet, two octets and one singlet for the group $SU(3)_f$. This is just the decomposition in Eq.~\eqref{333}
    for $SU(3)_f$! 

    Adding charm as the fourth flavor, the procedure is easily extended to $SU(4)_f$ and yields
    \begin{equation}\label{444}
       \mathbf{4}\otimes\mathbf{4}\otimes\mathbf{4} = \mathbf{20}_S \oplus \mathbf{20}_{M_A} \oplus \mathbf{20}_{M_S} \oplus \mathbf{4}_A\,.
    \end{equation}
     The first four columns in Table~\ref{tab:flavor} contain the $SU(2)_f$ wave functions: 
    \begin{equation}\label{222}
       \mathbf{2}\otimes\mathbf{2}\otimes\mathbf{2} = \mathbf{4}_S \oplus \mathbf{2}_{M_A} \oplus \mathbf{2}_{M_S}\,.
    \end{equation}

    If we consider $SU(2)$ \textit{spin} instead of flavor (replace $u\to \,\uparrow$, $d\to\,\downarrow$), we may also read off the
    spin wave functions. There are four singlets for spin 3/2,
    \begin{equation}\label{spin-3/2}
       \mS_s \quad \propto \quad \uparrow\uparrow\uparrow\,, \quad
       \uparrow\uparrow\downarrow + \uparrow\downarrow\uparrow + \downarrow \uparrow\uparrow\,, \quad
       \uparrow\downarrow\downarrow + \downarrow \uparrow\downarrow + \downarrow\downarrow \uparrow\,, \quad
       \downarrow\downarrow\downarrow\,,
    \end{equation}
     with subscript $s$ for spin,
    and two doublets for spin 1/2: 
    \begin{equation}\label{spin-1/2}
        \mD_s \quad \propto \quad \left[ \uparrow \downarrow \uparrow - \downarrow \uparrow \uparrow \atop -\frac{1}{\sqrt{3}}\left( \uparrow \downarrow \uparrow + \downarrow\uparrow \uparrow - 2\uparrow\uparrow \downarrow\right) \right], \quad
        \left[ \uparrow \downarrow\downarrow - \downarrow \uparrow \downarrow \atop \frac{1}{\sqrt{3}}\left( \uparrow \downarrow\downarrow + \downarrow \uparrow \downarrow - 2\downarrow\downarrow\uparrow\right) \right].
    \end{equation}

    The fact that the $SU(3)_f$ flavor symmetry is broken due to the unequal quark masses implies that the states inside the multiplets are
    no longer mass-degenerate. Although $I_3$ and $Y$ are still conserved, the Casimirs of $SU(3)_f$, which distinguish the multiplets, are no longer good quantum numbers.
    Therefore, states with the same $I_3$ and $Y$ can mix. In practice, the isospin symmetry $SU(2)_f$ is still approximately realized through $m_u \approx m_d$,
    which means that only states with the same $I$, $I_3$ and $Y$ will mix. This leads to an octet-decuplet mixing for the $\Sigma$ and $\Xi$ states and a singlet-octet mixing
    for the $\Lambda$ states.

\section{Quark models} \label{sec:qm}

    The statements inferred so far have been inferred from QCD's symmetries alone. What can we learn from the dynamics?
    Suppose we write the total wave function $\Psi$ of a baryon as
    \begin{equation}\label{wf}
       \Psi  = \Psi_\text{dynamics} \otimes \Psi_\text{flavor} \otimes \Psi_\text{color}\,.
    \end{equation}
    The color part is totally antisymmetric ($\Psi_\text{color} \propto \mA$) by Eq.~\eqref{333}.
    The flavor part can form singlets ($\mS_f$), doublets ($\mD_f$) and antisinglets ($\mA_f$), with $f$ for flavor,
    as given in Table~\ref{tab:flavor}.  What about the rest?

\begin{figure*}
      \centering
      \includegraphics[width=0.5\textwidth]{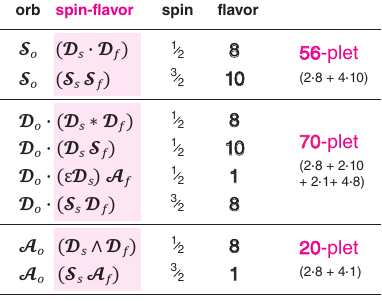}
      \caption{Spin-flavor wave functions in the SU(6)-symmetric quark model.}
      \label{fig:flavor-wfs}
    \end{figure*}

    Suppose we separate the dynamical part into $\Psi_\text{orbital} \otimes \Psi_\text{spin}$,
    where $\Psi_\text{orbital}$ denotes the spatial wave function and
    $\Psi_\text{spin}$ are the $SU(2)$ spin wave functions $\mS_s$ and $\mD_s$ from Eqs.~(\ref{spin-3/2}--\ref{spin-1/2}).
    Let us assume that the full wave function is totally antisymmetric due to the Pauli principle, i.e., an antisinglet $\mA$.
    Then, having arranged color, flavor and spin into permutation-group multiplets, this also fixes the symmetry of the orbital part.
    The resulting combinations are collected in Fig.~\ref{fig:flavor-wfs}, where $\mS_o$, $\mD_o$ and $\mA_o$ denote the orbital wave functions.
    The spin-flavor states exhaust all possible group-theoretical combinations\footnote{With
    $\mD = \left[a \atop s\right]$, $\mD' = \left[a' \atop s'\right]$, the dot, wedge and star products in Fig.~\ref{fig:flavor-wfs} are defined by~\cite{Eichmann:2015nra}
    \begin{equation}
       \begin{array}{rl}
           \mD \cdot \mD' &= aa' + ss'\,, \\[1mm]
           \mD \wedge \mD' &= as' - sa'\,,
       \end{array} \qquad
       \mD \ast \mD' = \left[ as' + sa' \atop aa' - ss' \right], \qquad
       \varepsilon = \left( \begin{array}{r @{\;\;\;}r} 0 & 1 \\ -1 & 0 \end{array}\right).
    \end{equation}
    The product $\mD\cdot\mD'$ is a singlet, $\mD\wedge\mD'$ an antisinglet, and $\mD\ast\mD'$ and ($\varepsilon\mD)\mA$ are doublets.}
    and yield 56 singlets, 70 doublets and 20 antisinglets.
    This is the `$SU(6)$-symmetric quark model' with $\mathbf{6}\otimes\mathbf{6}\otimes\mathbf{6} = \mathbf{56}_S \oplus \mathbf{70}_{M_A} \oplus \mathbf{70}_{M_S} \oplus \mathbf{20}_A$,
    where each elementary $\mathbf{6}$ is the combination of two spin and three flavor states.
    In this way, the spin, flavor and color parts are fully determined by their symmetries and only the spatial wave functions remain to be
    computed dynamically.

    A straightforward way to do so is to employ quark potential models.
    In nonrelativistic quantum mechanics one starts from the Schrödinger equation
    \begin{equation}\label{SGE}
       H\Psi = E\Psi\,, \qquad H = H_0 + \sum_{i<j} V(\vect{r}_{ij}) \,,
    \end{equation}
    where the potential is the sum of pairwise interactions between the quarks.
    An influential quark model is based on the Breit-Fermi interaction as the nonrelativistic limit of a one-gluon exchange~\cite{DeRujula:1975qlm},
    \begin{equation}
       V(\vect{r}) \propto \alpha \left[ -\frac{1}{r} + V_\text{ss}(\vect{r}) + V_\text{so}(\vect{r}) + \dots\right].
    \end{equation}
    It consists of a Coulomb part, a hyperfine (spin-spin) interaction that generates the splittings between $S=\tfrac{1}{2}$ and $\tfrac{3}{2}$ states like the $N-\Delta$ splitting, and
    further spin-orbit terms. A long-range confinement potential $V_\text{conf} \propto r$ can then be added by hand.
    There is a wide spectrum of nonrelativistic and relativistic quark models, see e.g.~\cite{Hendry:1978ee,Capstick:2000qj,Klempt:2009pi,Richard:2012xw,Crede:2013sze,Thiel:2022xtb} for reviews,
    including quark models
    based on one-gluon exchange~\cite{DeRujula:1975qlm,Isgur:1979be,Capstick:1986bm}, Goldstone-boson exchange~\cite{Glozman:1995fu,Melde:2008yr},   
    diquark models~\cite{Anselmino:1992vg,Ebert:2007nw,Santopinto:2014opa}, string models~\cite{Bijker:1994yr},
    large-$N_c$ models~\cite{Goity:2003ab,Matagne:2004pm}, hypercentral quark models~\cite{Giannini:2001kb,Giannini:2015zia}, light-front and holographic models~\cite{Brodsky:2014yha}, 
    and  models based on Bethe-Salpeter equations~\cite{Loring:2001kv,Metsch:2008zz}.

\begin{figure*}
      \centering
      \includegraphics[width=1\textwidth]{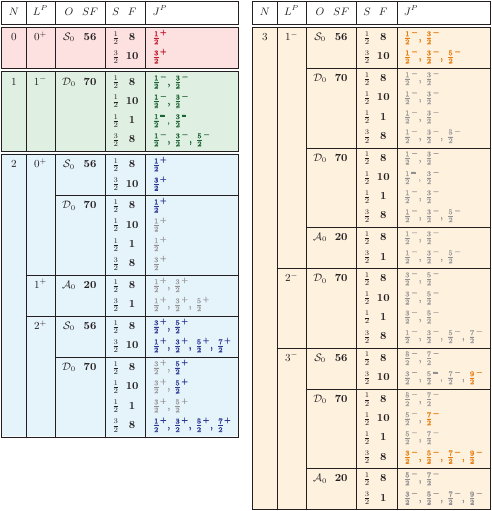}
      \caption{Baryon spectrum in the harmonic oscillator model with band quantum number $N$,  spin $S$, orbital angular momentum $L$, total angular momentum $J$ and parity $P$.
               $O$ denotes the symmetry of the orbital part and $SF$ the spin-flavor multiplets. The $J^P$ quantum numbers in bold color can be identified with well-established (two-, three-, four-star) states~\cite{Crede:2013sze,Thiel:2022xtb},
               whereas those in gray do not have good experimental candidates.}
      \label{fig:missing-resonances}
      \vspace{-5mm}
    \end{figure*}

    In practice, the spatial wave functions are often set up in a spherical
    harmonic oscillator basis~\cite{Klempt:2009pi,Klempt:2012fy,Crede:2013sze}. After removing the center-of-mass motion, they depend on two Jacobi variables
    $\vect{\rho} = (\vect{x}_1 - \vect{x}_2)/\sqrt{2}$ and $\vect{\lambda} = (\vect{x}_1 + \vect{x}_2 -2\vect{x}_3)/\sqrt{6}$,
    \begin{equation}
       \phi_L(\vect{\rho},\vect{\lambda}) = \sum_{n_\rho,l_\rho, n_\lambda,l_\lambda} c_{n_\rho l_\rho n_\lambda l_\lambda}^L \left[ \phi_{n_\rho l_\rho}(\vect{\rho})\otimes \phi_{n_\lambda l_\lambda}(\vect{\lambda})\right]_L\,,
    \end{equation}
    which allows for radial ($n_\alpha > 0$) and orbital excitations ($l_\alpha > 0$). With $n=n_\rho + n_\lambda$ and $l=l_\rho + l_\lambda$
    one arrives at the `band quantum number' $N = 2n+l$, where for a pure oscillator potential
    the states with the same $N$ would carry the same energy.
    The quantum numbers $J^P$ are then obtained from the parity $P=(-1)^l$, the quark spin $S = \tfrac{1}{2}$, $\tfrac{3}{2}$ and the orbital angular momentum
    $L = |l_\rho - l_\lambda| \dots l_\rho+ l_\lambda$.
    The resulting states are collected in Fig.~\ref{fig:missing-resonances}.
    The harmonic-oscillator levels $(N, L^P, SF) = (0, 0^+, \mathbf{56})$ can be identified with the nucleon
    and $\Delta(1232)$ ground states;
    the $(1, 1^-, \mathbf{70})$ orbital excitations with the first few negative-parity states such as the $N(1535)$, $N(1520)$ etc.;
    the $(2, 0^+, \mathbf{56})$ radial excitations with the Roper resonance $N(1440)$ and $\Delta(1600)$, and so on.

        As one can see from Fig.~\ref{fig:missing-resonances}, this predicts \textit{a lot} of states.
        While the bands for $N=0$ and $N=1$ can be identified with experimentally known baryons,
        already the $N=2$ and especially the $N=3$ states have not all been observed.
        This is the so-called \textit{missing resonances problem}, which could have several (not mutually exclusive) explanations:

       \smallskip
           {\tiny$\blacksquare$} We simply have not  found them yet. Excited $N$ and $\Delta$ baryons
              have traditionally been extracted from $N\pi$ scattering, 
              but if they did not strongly couple to $N\pi$
              it would be hard to see their peaks in experimental cross sections.
              New photoproduction experiments and improved partial-wave analyses have indeed added several new states to the PDG~\cite{Thiel:2022xtb}, 
              but the spectrum as of today (Fig.~\ref{fig:spectrum-1}) is still quite sparse compared to what the quark model predicts.

       \smallskip
           {\tiny$\blacksquare$} If two quarks inside a baryon clustered to a \textit{diquark},  this would freeze internal excitation degrees of freedom
              and we might see fewer states in the spectrum. While the simplest possibility of \textit{pointlike} diquarks  
              is  disfavored by comparison with experiment~\cite{Nikonov:2007br},
              functional calculations in QFT do support strong quark-quark correlations in baryons, as we will discuss in Sec.~\ref{sec:fm}.

       \smallskip
           {\tiny$\blacksquare$} Light baryons could have substantial \textit{multiquark} admixtures such as $(qqq)(q\bar{q})$,  $(qqq)(q\bar{q})(q\bar{q})$ etc.,
           which ties into the question of meson-cloud effects and dynamically generated resonances.
       Among the prime candidates are the Roper resonance and the $\Lambda(1405)$, see e.g.~\cite{Burkert:2017djo,Meissner:2020khl,Mai:2020ltx}. 

       \smallskip
           {\tiny$\blacksquare$} The assumptions we made (nonrelativistic quark model, harmonic oscillator)
              may just be too drastic  to provide a realistic description of light baryons.
     The Poincaré group has two Casimirs $M$ and $J$, whereas the quark spin $S$ and orbital angular momentum $L$
    can mix in different frames, so they are not good quantum numbers  to label the states. In QFT we also lose the  concept
    of a `wave function' with a probability interpretation ---  its role is instead played by the \textit{Bethe-Salpeter}  wave function, cf.~Eq.~\eqref{bswf} below. It
    can still be cast in the form~\eqref{wf} but  involves many more components
    and relativistically there will be new effects. As we will see in Sec.~\ref{sec:fm}, a state with $J^P = 1/2^+$ like the nucleon
    can carry $L=0,1,2$, where the $L=1$ components (the `$p$ waves') contribute a substantial amount to the wave function.

\section{Quantum field theory toolbox} \label{sec:qft}

    While nonrelativistic quark models are well-suited to tackle the heavy-quark spectrum,
    relativity and chiral symmetry complicate the dynamics in the light-quark sector. 
    Unfortunately, translating the Schrödinger equation~\eqref{SGE} to relativistic QFT is not straightforward.
    On the one hand, relativity forces us to give up the concept of a wave function with a probability interpretation.
    On the other hand, a QFT describes particle creation and annihilation such that the Hilbert space becomes infinite-dimensional,
    and one has to deal with renormalization which is non-trivial in a Hamiltonian description.
    While some of these issues are addressed in light-front Hamiltonian approaches~\cite{Brodsky:2014yha,Hiller:2016itl},
     baryon spectrum calculations in QFT often do not start from a Hamiltonian except in certain limits (like heavy quarks)
    or under (e.g. non-relativistic) approximations. Then how else?

    \begin{figure*}
      \centering
      \includegraphics[width=1\textwidth]{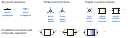}
      \caption{Some of the lowest $n$-point correlation functions in QCD.}
      \label{fig:cfs}
    \end{figure*}

    Let us briefly recap the main building blocks of a QFT. The starting point is the classical action $S[\phi] = \int d^4x \,\mL(\phi)$,
    which encodes the  fields $\phi_i$ and their  interactions. In QCD,  $\phi_i = \{ \psi, \conjg\psi, A^\mu\}$  are the quark, antiquark and gluon fields.
    In the QFT, the fields $\phi_i(x)$ become operators on a state space consisting of the vacuum $|0\rangle$,
    one-particle states $|p\rangle$ with an onshell momentum $p$, and multiparticle states $|p_1 \dots p_n\rangle$.
    Relativity and unitarity demand that these states transform under unitary representations of the Poincaré group, which ensures a probability interpretation for S-matrix elements
    $\langle p_1 \dots p_n | q_1 \dots q_n\rangle$. Furthermore, causality tells us that two measurements at spacelike distances cannot affect each other,
    which implies (anti)\,commutation relations for bosonic (fermionic) fields.
    These are the basic pillars of a QFT,
    and one can extend the list by adding the spectral condition (for physical states), renormalizability (for renormalizable theories),
    or gauge invariance (for gauge theories).

    Unfortunately, for an interacting QFT one quickly runs into mathematical troubles at the operator level. However, the \textit{n-point correlation functions}
    \begin{equation}\label{gf}
       G(x_1, \dots x_n) = \langle 0 | \mathsf{T}\,\phi(x_1) \dots \phi(x_n) | 0 \rangle
    \end{equation}
    of the theory, where $\mathsf{T}$ denotes time-ordering, are still well-defined. Therefore, they become the central objects in a QFT:
    The set of all (infinitely many) $n$-point functions contains the full content of the QFT,
    and they are related to onshell S-matrix elements which can be measured experimentally.
    Thus, a central task in a QFT is to compute its $n$-point functions.

    The lowest $n$-point functions of QCD are illustrated in Fig.~\ref{fig:cfs}: These are the two-point functions such as the quark and gluon propagators,
    the three-point functions (quark-gluon vertex, three-gluon vertex), as well as higher $n$-point functions such as the four-gluon vertex,
    the quark four-point function, the quark six-point function, etc. One can also construct correlation functions with composite operators, e.g.
    of the form $\langle 0 | (\conjg\psi \Gamma \psi)(x)\,(\conjg\psi \Gamma' \psi)(y) |0 \rangle$ with some Dirac-flavor-color matrices  $\Gamma$, $\Gamma'$;
    these are the central objects of interest in lattice QCD calculations.

    In Quantum Electrodynamics (QED), due to the smallness of the electromagnetic coupling, the respective $n$-point functions can be computed in perturbation theory
    by expanding it into Feynman diagrams. Because they are also directly connected to physical S-matrix elements (such as $e^- e^-$ or $e^+ e^-$ scattering),
    this has led to extremely precise QED predictions, e.g. for the anomalous magnetic moment of the electron and muon~\cite{Aoyama:2020ynm}.
    On the other hand, a perturbative expansion in QCD is only useful at high energies, where the strong coupling is small,
    whereas at low energies it becomes large and  perturbation theory is no longer applicable. Moreover, in QCD the \textit{physical} onshell states
    are \textit{bound states}, namely hadrons, and thus there is no longer a 1:1 correspondence between S-matrix elements, whose external legs are hadrons,
    and the elementary $n$-point functions made of quarks and gluons.

    \begin{figure*}
      \centering
      \includegraphics[width=1\textwidth]{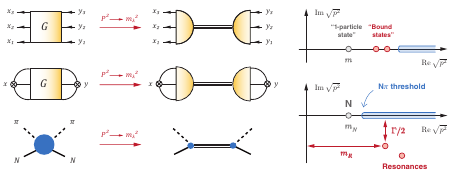}
      \caption{Baryon poles appear in any correlation function that can produce them, such as the quark six-point function,
               the respective current correlator or the $N\pi$ scattering amplitude.}
      \label{fig:poles}
    \end{figure*}

    In turn, an important relation in QFT  says that physical states produce \textit{poles} in correlation functions~\cite{Weinberg:1995mt}.
    If we split the $n$ coordinates in Eq.~\eqref{gf} into two sets $\{x_i\}$ and $\{x_j\}$ with $i=1\dots N$ and $j=1\dots N'$, then by inserting the completeness relation of the state space one can show that
    each onshell state $|\lambda\rangle$ with momentum $P$ and mass $m_\lambda$ produces a pole in any correlation function that is compatible with the quantum numbers of the state:
    \begin{equation}\label{poles}
       G(x_1, \dots x_n) = \int \!  \frac{d^4P}{(2\pi)^4}\,e^{-iPz}\left[ \frac{i\Psi(\{x_i\},P)\,\Psi^\dag(\{x_j\},P)}{P^2 - m_\lambda^2 + i\epsilon} + \text{finite}\right].
    \end{equation}
    The residue at the pole defines the Bethe-Salpeter wave function (BSWF),
    \begin{equation}\label{bswf}
       \Psi(\{x_i\},P) = \langle 0 | \mathsf{T}\,\phi(x_1) \dots \phi(x_N) | \lambda \rangle\,,
    \end{equation}
    where an overall coordinate $z$ has been factored out by translation invariance. 
    $\Psi$ can be viewed as the QFT analogue of the quantum-mechanical wave function since it carries the information about the hadron, except it does not have a direct probability interpretation.
    Applied to QCD, the one-particle states $|\lambda\rangle$ are stable hadrons, and therefore such poles appear in any channel that can create gauge-invariant hadron quantum numbers.

    Eq.~\eqref{poles} is the central formula that allows one to extract baryon properties from QCD, as illustrated in Fig.~\ref{fig:poles}.
    It applies equally to the (gauge-dependent) elementary quark six-point function, the (gauge-invariant) two-point current correlators
    that arise from using composite operators with contracted quark fields, as well as any scattering amplitude that admits baryon poles
    such as e.g. $N\pi$ scattering. The physical singularity locations are independent of the correlation function: In the simplest case, a two-point function
    can produce one-particle poles, bound-state poles, and branch cuts that signal the onset of the multiparticle continuum.
    For an $NN$ correlator, the `one-particle state' is the nucleon itself and the lowest multiparticle threshold is $N\pi$. The branch cut
    is pulled over the `bound-state' poles which can turn into resonances in the complex plane on a higher Riemann sheet,
    whose real and imaginary parts describe the mass and the width of the resonance.

    It is interesting to note that the machinery of QFT does not actually tell us \textit{what} a hadron is. From Fig.~\ref{fig:poles},
    a nucleon resonance can be thought of as an $N\pi$ state, but also as a three-quark state ($qqq$), or a five-quark state ($qqqq\bar{q}$) etc.,
    because all those correlation function must produce the same gauge-invariant spectrum. All we can say is how much \textit{overlap} a state has
    with a given configuration, and this overlap is defined by the respective BSWF.

    So how does one extract the baryon spectrum in practice?

    \smallskip
    {\tiny$\blacksquare$}
    Effective field theories start from the third option in Fig.~\ref{fig:poles}, namely
    by employing multichannel scattering equations such as for $N\pi\to N\pi$, $N\gamma^{(\ast)}\to N\pi$, etc. 
    Because such approaches start from hadronic Lagrangians, the internal hadrons in the loop diagrams are in principle offshell,
    and one may distinguish `bare' vs. `dressed' states to quantify the effect of meson-baryon interactions.
    Multichannel amplitude analyses have  been instrumental in the experimental extraction of the baryon spectrum, see~\cite{Thiel:2022xtb,Anisovich:2016vzt,Kamano:2019gtm} and references therein.  

    \smallskip
    {\tiny$\blacksquare$}
    The second row in Fig.~\ref{fig:poles} is the basis of lattice QCD calculations, where  gauge-invariant current correlators
    are computed through the QCD path integral by statistically sampling over the quark and gluon field configurations.
    The nucleon excitation spectrum is contained, for example, in\footnote{From
    now on we will use a Euclidean metric (+,+,+,+), see e.g. Appendix A of Ref.~\cite{Eichmann:2016yit} for conventions.
    For our purposes, this implies $p^2 \to -p^2$, $\slashed{p} \to -i\slashed{p}$ and $e^{iS} \to e^{-S}$.}
    \begin{equation}\label{path-int}
       G(x-y) = \langle 0 | \mathsf{T}\,N(x)  \bar N(y) | 0 \rangle \propto \int \mathcal{D}[\psi,\bar\psi,A]\,e^{-S[\psi,\bar\psi,A]}\,N(x) \bar{N}(y)\,,
    \end{equation}
    where $N(x)$ is some contraction of three quark field operators.
    At large Euclidean times, the correlator falls off exponentially, $G(\tau\to \infty) \propto e^{-m\tau}$,
    which translates to a pole with mass $m$ in momentum space. Lattice calculations are done in a finite volume,
    which does not give direct access to the singularity structure in the complex plane, but there are efforts
    to extract the resonance spectrum using generalizations of the Luescher method, see e.g.~\cite{Briceno:2017max,Mai:2018djl,Mikhasenko:2019vhk,Hansen:2020otl}.
    From Figs.~\ref{fig:cfs}--\ref{fig:poles}, employing different lattice operators translates to contracting different $n$-point functions,
    e.g. also those with gluon legs, and lattice calculations indicate that a number of excited baryons have large gluon components~\cite{Edwards:2011jj,Dudek:2012ag}.

    \smallskip
    {\tiny$\blacksquare$}
    Finally, the first row in Fig.~\ref{fig:poles} is the starting point for functional methods, where
    the strategy is to calculate the quark and gluon $n$-point correlation functions directly and  extract the physical spectrum afterwards.
    This will be our focus in the remaining section.

         \begin{figure*}[t]
      \centering
      \includegraphics[width=1\textwidth]{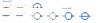}
      \caption{Dyson-Schwinger equations for the quark and gluon propagator.}
      \label{fig:dses}
    \end{figure*}

\section{Functional methods} \label{sec:fm}

   The jumping-off point for functional methods is again the path integral in Eq.~\eqref{path-int}. However,
   instead of calculating the correlation functions directly, one derives relations that couple them to each other.
   This leads to different systems of self-consistent  equations like the Dyson-Schwinger equations (DSEs)~\cite{Roberts:1994dr,Alkofer:2000wg,Huber:2018ned} and
   functional renormalization-group equations~\cite{Pawlowski:2005xe,Dupuis:2020fhh}.

   Because the resulting equations are infinitely coupled, in practice one needs to employ truncations.
   This is illustrated in Fig.~\ref{fig:dses}: The quark DSE is a self-consistent integral equation and determines the inverse quark propagator $S(p)^{-1}$,
   i.e., its two dressing functions $A(p^2)$ and $M(p^2)$,
   \begin{equation}\label{qprop}
      S(p)^{-1} = A(p^2)\left(i\slashed{p} + M(p^2)\right),
   \end{equation}
   where $M(p^2)$ is the quark mass function.
   The necessary input are the gluon propagator and quark-gluon vertex. 
   Now, we could  employ ansätze for them and solve just the quark DSE. Or we could also solve
   the gluon DSE, which depends again on the quark propagator and quark-gluon vertex, but also on the three- and four-gluon vertex, etc.
   We could employ ansätze for those, or we could solve also their  DSEs which depend on higher $n$-point functions, and so on.
   By systematically enlarging the set of computed $n$-point functions in this way, one should eventually get more and more precise results,
   and to some extent this is  already seen in the existing calculations~\cite{Cyrol:2017ewj,Huber:2020keu,Eichmann:2021zuv,Pawlowski:2022oyq}.

        \begin{figure*}
      \centering
      \includegraphics[width=1\textwidth]{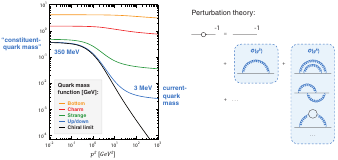}
      \caption{Quark mass function $M(p^2)$ for different flavors from Dyson-Schwinger calculations~\cite{Eichmann:2016yit}.
               In the chiral limit, the perturbative expansion leads to $M(p^2)=0$.}
      \label{fig:quark-dse}
    \end{figure*}

   The $n$-point functions are not merely a means to do spectroscopy but also interesting in their own right. 
   The gluon propagator carries information on confinement and gluon mass generation~\cite{Aguilar:2016ock,Cyrol:2016tym,Eichmann:2021zuv,Aguilar:2021uwa}.
   The quark propagator encodes the dynamical generation of a quark mass at low momenta, which is shown in Fig.~\ref{fig:quark-dse}.
   At large momenta, $M(p^2) = m_f$ is the current-quark mass that enters in the QCD Lagrangian. At low momenta
   a large `constituent-quark' mass is dynamically generated, which is about 350 MeV for all quark flavors,
   so it is washed out for heavier quarks. 

   The origin of this behavior is the spontaneous breaking of chiral symmetry
   due to the quark-gluon dynamics. This is a nonperturbative effect:
   By expanding the quark DSE in perturbation theory, which is justified for large momenta where the coupling is small, 
   one successively generates all  Feynman diagrams for the propagator (Fig.~\ref{fig:quark-dse}). In the chiral limit ($m_f = 0$), each perturbative propagator is proportional to $\slashed{p}$
   and each vertex to $\gamma^\mu$. Now, each diagram in Fig.~\ref{fig:quark-dse} contains an odd number of $\gamma$ matrices, whose trace vanishes. According to Eq.~\eqref{qprop}, the trace
   is proportional to the mass function $M(p^2)$, which must therefore vanish to all orders in perturbation theory.
   Thus, a non-zero mass function must be a nonperturbative effect.
   Because chiral symmetry demands $M(p^2) = 0$, this also implies that chiral symmetry is spontaneously broken.
   In fact, the dynamical generation of a quark mass can already be illustrated  with analytically solvable models~\cite{Roberts:2007jh,Eichmann:2016yit}.
   This relates to the question from Sec.~\ref{sec:intro}: Why is the proton mass not simply the sum of its three current-quark masses? Because QCD generates the rest!

\begin{figure*}
      \centering
      \includegraphics[width=1\textwidth]{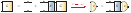}
      \caption{Scattering equation and homogeneous Bethe-Salpeter equation at the pole.}
      \label{fig:bse}
    \end{figure*}

   So what about hadrons? According to Fig.~\ref{fig:poles}, we should push the calculation of $n$-point functions to high orders
   to  eventually inclulde the quark six-point function, from where we can extract baryon poles.
   Unfortunately this  would become quite cumbersome due to the complicated structure and equations for these objects.
   A more efficient strategy is to solve Bethe-Salpeter equations (BSEs), like the one shown in Fig.~\ref{fig:bse} for the $q\bar{q}$ four-point function.
   The latter satisfies a scattering equation, which reads in a compact notation
   \begin{equation}\label{sceq}
      G = G_0 + G_0\,K\,G\,.
   \end{equation}
   This is again an integral equation, where every multiplication stands for a four-momentum integration.
   Upon iteration, the equation generates all possible diagrams that contribute to $G$.
   $K$ is the $q\bar{q}$ irreducible kernel and contains all diagrams that do not fall apart by cutting two horizontal quark lines, because those are already generated by the iteration.

   The fact that the equation is \textit{nonperturbative} can be seen by the  analogy with the geometric series: the result of the equation
   \begin{equation}
      f(x) = 1 + x f(x) = 1 + x + x^2 f(x) = 1 + x + x^2 + x^3 f(x) = \dots
   \end{equation}
   is $f(x) = 1/(1-x)$,
   whereas the series $f(x) = 1 + x + x^2 + x^3 + \dots$  converges to that result only if $|x| < 1$.
   Similarly, for a small coupling Eq.~\eqref{sceq} can be expanded in the perturbative series $G = G_0 + G_0\,K\,G_0 + G_0\,K\,G_0\,K\,G_0 + \dots $
   This will, however, not produce hadron poles.
   The nonperturbative result, on the other hand, has the formal structure $G^{-1} = G_0^{-1} - K$ and will produce poles if `$KG_0=1$' at some $P^2 = -m^2$.

   By comparing the residues of Eq.~\eqref{sceq} at a given pole location,  one arrives at the homogeneous BSE
   for the BSWF:
   \begin{equation}
      \Psi = G_0 K \,\Psi \quad \Leftrightarrow \quad  KG_0\,\Gamma_i = \lambda_i\,\Gamma_i\,.
   \end{equation}
   It is customary to consider $\Psi$ or the corresponding Bethe-Salpeter amplitude $\Gamma$ defined by removing the propagator legs, $\Psi = G_0 \Gamma$,
   in which case the BSE becomes $\Gamma = K G_0 \Gamma$. In practice the equation is solved by computing the eigenvalue spectrum of $K G_0$ for different values of $P^2 = -m^2$.
   If an eigenvalue satisfies $\lambda_i(P^2 = -m_i^2) = 1$, one has found a ground or excited state of the system, 
   in which case $G$ has a pole. 
   Note that this also applies to resonances with poles in the complex $P^2$ plane on higher Riemann sheets
   since nothing forces $P^2$ to be real.

   The remaining question is what the kernel $K$ looks like. Much of the existing work on meson spectroscopy has been done in a rainbow-ladder truncation~\cite{Maris:1997tm,Maris:1999nt,Maris:2005tt,Hilger:2014nma,Rojas:2014aka,Fischer:2014xha}.
   Here the kernel is assumed to be a gluon exchange between the quark and antiquark, and one only keeps the classical tensor $\gamma^\mu$ in the quark-gluon vertex.
   This preserves chiral symmetry and its dynamical breaking pattern:
   when solving the quark DSE, one dynamically generates quark masses like in Fig.~\ref{fig:quark-dse},
   and the pseudoscalar mesons calculated from their BSEs are the massless Goldstone bosons in the chiral limit.
   In the light-meson sector, rainbow-ladder works quite well for pseudoscalar and vector mesons, whereas scalar and axialvector mesons are too strongly bound.
   There are ongoing efforts to go beyond rainbow-ladder by also implementing the information from the gluon sector and higher $n$-point functions,
   and these have already considerably improved the meson spectrum~\cite{Chang:2011ei,Williams:2015cvx,Huber:2020ngt}.

\begin{figure*}
      \centering
      \includegraphics[width=1\textwidth]{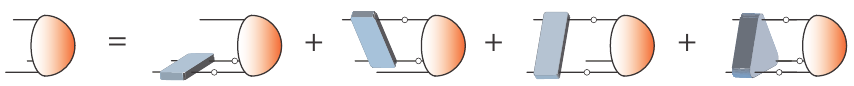}
      \caption{Three-body Bethe-Salpeter or Faddeev equation.}
      \label{fig:qqq}
    \end{figure*}

   The analogous three-body BSE is shown in Fig.~\ref{fig:qqq}, where the kernel is the sum of irreducible $qq$ and $qqq$ components.
   It is interesting to note that the simplest diagram in the $qqq$ kernel, namely the three-gluon vertex connecting the three quarks,
   vanishes simply by the color algebra. This is a first hint that baryons may be dominated by \textit{two-quark} correlations.

   The earlier statement that the BSWF can be viewed as the generalization of the baryon's wave function in quantum mechanics
    becomes clear when we write down its full structure according to Poincaré invariance:
   \begin{equation*}\label{baryon-amp}
     \Psi_{\alpha\beta\gamma\delta}(p,q,P) = \sum_{i} f_i(p^2, q^2, p\cdot q, p\cdot P, q\cdot P)\,\tau_i(p,q,P)_{\alpha\beta\gamma\delta} \otimes \text{Flavor} \otimes \text{Color}\,.
   \end{equation*}
   The flavor and color parts are identical to Eq.~\eqref{wf},
   but $\Psi_\text{dynamics}$ is more general. 
   It has three Dirac indices for the quarks and one for the nucleon,
   and it depends on three momenta $p$, $q$, $P$, which makes six Lorentz invariants, one of them being onshell ($P^2 = -m^2$).
   As a consequence, $J=1/2$ states depend on 64 linearly independent Dirac tensors $\tau_i$ and $J=3/2$ states on 128~\cite{Eichmann:2009qa,Sanchis-Alepuz:2011egq}.
   Their dressing functions $f_i$ carry the dynamical information.

   One can still arrange these tensors in eigenstates of the spin $S$ and orbital angular momentum $L$ in the baryon's rest frame,
   in order to obtain tensors with $L=0$ ($s$ waves), $L=1$ ($p$ waves), $L=2$ ($d$ waves), and so on (cf. Table~2 in Ref.~\cite{Eichmann:2009en}). In general, however, all these components
   mix together in the BSWF to produce a baryon with definite $J^P$.
   For example, for the proton the $f_i \,\tau_i$ can be rearranged in permutation-group doublets to arrive at a fully antisymmetric wave function $\sum_{i=1}^{64} \mD_i\cdot \mD_f\,\mA_c$,
   but in contrast to the first row in Fig.~\ref{fig:flavor-wfs} this does not just give two components (spin up and down) but 64.
   Moreover, many of those belong to $L=1$, so that relativistically even the nucleon carries a large fraction of $p$ waves (cf.~Fig.~\ref{fig:spectrum-qdq} below).

   A convenient feature of the three-body BSE in Fig.~\ref{fig:qqq} is that, after dropping the irreducible three-quark part,
   one can take over the rainbow-ladder kernel from the meson sector without any adjustments. The interaction models that are frequently employed~\cite{Maris:1999nt,Qin:2011dd}
   depend  on two parameters, a scale and a shape parameter. The shape parameter does not affect a wide range of observables including baryon ground-state masses,
   so  the scale remains the only active parameter. If it is adjusted to reproduce the experimental pion decay constant, the nucleon mass from the
   three-body equation comes out at $m_N = 0.94$ GeV~\cite{Eichmann:2011vu}. Furthermore, also the ground-state masses for hyperons and heavy quarks agree very well with experiment~\cite{Eichmann:2016yit,Sanchis-Alepuz:2014sca,Qin:2019hgk}.

        \begin{figure*}
      \centering
      \includegraphics[width=1\textwidth]{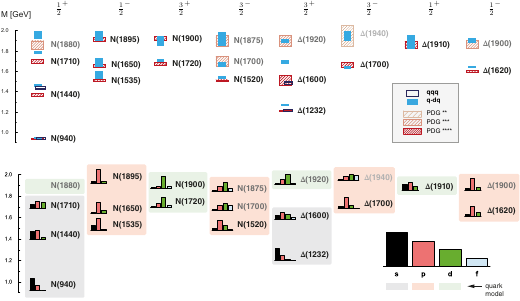}
      \caption{Top: Light baryon spectrum from the three-quark and quark-diquark calculations for different $J^P$ quantum numbers~\cite{Eichmann:2016hgl,Eichmann:2016nsu}, 
               compared to the PDG masses (real parts of pole positions)~\cite{ParticleDataGroup:2020ssz}.
               Bottom: Orbital angular momentum contributions to each baryon from the quark-diquark calculation. The bars sum up to 100\%, and the light-colored boxes drawn around them
               are the quark-model values for $L$ from Fig.~\ref{fig:missing-resonances}.}
      \label{fig:spectrum-qdq}
    \end{figure*}

   Because of the rich structure of the baryon wave functions, 
   the calculation of excited baryons amounts to eigenvalue equations for very large matrices which are computationally  expensive.
   A more efficient strategy is to employ a \textit{quark-diquark} description:
   With  a few assumptions one can convert the three-body equation to a quark-diquark BSE~\cite{Eichmann:2016yit,Barabanov:2020jvn,Oettel:1998bk,Roberts:2011cf}. 
   Here the gluons no longer appear and the interaction proceeds through a ping-pong quark exchange between the other two quarks,
   thus forming (non-pointlike) diquarks in the process. In turn, the diquark ingredients are calculated from their own equations using the same quark-gluon interaction.
   It turns out that the lowest-lying diquarks are the parity partners of the lowest-lying mesons~\cite{Maris:2002yu}:
   \begin{itemize}
   \item pseudoscalar mesons correspond to scalar diquarks (with $m_\text{sc}\sim 0.8$ GeV),
   \item vector mesons to axialvector diquarks ($m_\text{av}\sim 1$ GeV),
   \item scalar mesons to pseudoscalar diquarks,
   \item axialvector mesons to vector diquarks, and so on.
   \end{itemize}

   Now, the observation that scalar and axialvector mesons are too strongly bound in rainbow-ladder also translates to their diquark partners
   and the resulting baryons. These are the `orbital excitations' from the quark model, i.e., nucleons with $J^P = 1/2^-$, $3/2^\pm$ and $\Delta$ baryons with $J^P = 3/2^-$, $1/2^\pm$,
   which come out too light compared to experiment.
   This happens both in the three-quark and quark-diquark approach, even though the former does not know anything about diquarks -- the baryon spectrum is very similar in both cases~\cite{Eichmann:2016hgl}.
   Moreover, if one adjusts the strength in the higher-lying diquark channels to the meson spectrum in order to mimic effects beyond rainbow-ladder, 
   then one finds a 1:1 correspondence with experiment for the resulting baryon spectrum -- this is  shown in Fig.~\ref{fig:spectrum-qdq}.

   It is also interesting to study the composition of these baryons in terms of their orbital angular momentum contributions.
   The bars in Fig.~\ref{fig:spectrum-qdq} show the $s$, $p$, $d$ and $f$-wave contributions to each baryon's normalization.
   The light-colored boxes are the quark-model values for $L$ (according to Fig.~\ref{fig:missing-resonances}).
   In the majority of cases the nonrelativistic quark model does indeed shine through,
   e.g., the $N(1/2^-)$ states are dominated by $p$ waves and the $N(3/2^+)$ states by $d$ waves.
   But there are also clear deviations:
   the nucleon and $\Delta(1232)$ are dominated by $s$ waves but have substantial (relativistic) $p$-wave contributions;
   and for some cases like the Roper resonance, the $\Delta(1600)$, and the $\Delta(1910)$, the nonrelativistic assignment does not work at all.

   Taken together, these observations point towards the last option at the end of Sec.~\ref{sec:qm}:
   Relativity is clearly important for light baryons
   and can lead to entirely new effects for the spectrum.
   In fact, even though the relativistic contributions decrease for heavy baryons, their effects persist even up to bottom baryons~\cite{Qin:2018dqp}.
   Secondly, the picture of a diquark dominance in baryons is justified because the consistent treatment of the three-quark and quark-diquark approach  leads to a very similar spectrum.
   In any case, this does not rule out  other options: the admixture of multiquark components, which are not yet taken into account in these calculations, is most likely important,
   and it is safe to assume that many  baryon resonances still await to be found.

   There is a body of results on light baryons from functional methods also beyond spectroscopy: electromagnetic elastic and transition form factors~\cite{Eichmann:2011vu,Eichmann:2011aa,Segovia:2014aza,Segovia:2015hra},
   axial form factors~\cite{Eichmann:2011pv,Chen:2021guo} and distribution amplitudes~\cite{Mezrag:2017znp,Bednar:2018htv}, to name some of them.
   A major future goal is to go towards ab-initio calculations by computing also
   higher $n$-point functions and implement them in baryon calculations.
   Beyond rainbow-ladder studies have already improved the light meson spectrum and similar advances can be expected in the baryon sector.
   A related line of improvements is the investigation of the resonance structure and multiquark admixture, where first such calculations 
   have been explored in the meson sector~\cite{Williams:2018adr,Santowsky:2020pwd,Miramontes:2021xgn}.  

   \vspace{-5mm}

\section{Outlook}\label{summary}

   Baryon spectroscopy continues to be a cornerstone of strong interaction studies and a highly active area of research.
   A systematic description of baryon resonances will likely require combined efforts from
   experiment, lattice QCD, amplitude analyses, phenomenology, and functional methods.
   The current large interest in exotic hadron spectroscopy gives another strong motivation in 
   revisiting the `ordinary' baryon spectrum, which may very well be more exotic than we thought: relativity, chiral symmetry, multiquark admixtures and resonance dynamics
    make the study of baryons a fascinating topic to uncover the microscopic dynamics of QCD.

\begin{acknowledgements}
I would like to thank the organizers of the 2021  School on the Physics of Baryons for the invitation to present this lecture,
and I am grateful to Christian Fischer and Helios Sanchis-Alepuz who have been collaborating with me on these topics.
\end{acknowledgements}


\bibliographystyle{spmpsci_unsrt}

\bibliography{baryons}


\end{document}